\documentclass[conference]{IEEEtran}
\IEEEoverridecommandlockouts
\usepackage{cite}
\usepackage{amsmath,amssymb,amsfonts}
\usepackage{algorithmic}
\usepackage{graphicx}
\usepackage{subcaption}
\usepackage{colortbl}
\usepackage{textcomp}
\usepackage{xcolor}
\usepackage{colortbl}
\usepackage[normalem]{ulem}
\usepackage{silence}
\usepackage{array}
\newlength\mylength
\def\BibTeX{{\rm B\kern-.05em{\sc i\kern-.025em b}\kern-.08em
    T\kern-.1667em\lower.7ex\hbox{E}\kern-.125emX}}

\usepackage{tabularx}

\begin{document}

\title{Breast Cancer Diagnosis with Transfer Learning and Global Pooling}

\author{\IEEEauthorblockN{Sara Hosseinzadeh Kassani}
\IEEEauthorblockA{\textit{Department of Computer Science} \\
\textit{University of Saskatchewan}\\
Saskatoon, Canada \\
sara.kassani@usask.ca}
\and
\IEEEauthorblockN{Peyman Hosseinzadeh Kassani}
\IEEEauthorblockA{\textit{Department of Biomedical Engineering} \\
\textit{University of Tulane}\\
New Orleans, USA  \\
peymanhk@tulane.edu}
\and
\IEEEauthorblockN{Michal J. Wesolowski}
\IEEEauthorblockA{\textit{Department of Medical Imaging} \\
\textit{University of Saskatchewan}\\
Saskatoon, Canada \\
mike.wesolowski@usask.ca}
\and
\IEEEauthorblockN{Kevin A. Schneider}
\IEEEauthorblockA{\textit{Department of Computer Science} \\
\textit{University of Saskatchewan}\\
Saskatoon, Canada \\
kevin.schneider@usask.ca}

\and
\IEEEauthorblockN{Ralph Deters}
\IEEEauthorblockA{\textit{Department of Computer Science} \\
\textit{University of Saskatchewan}\\
Saskatoon, Canada \\
deters@cs.usask.ca}
}

\maketitle

\begin{abstract}
Breast cancer is one of the most common causes of cancer-related death in women worldwide. Early and accurate diagnosis of breast cancer may significantly increase the survival rate of patients. In this study, we aim to develop a fully automatic, deep learning-based, method using descriptor features extracted by Deep Convolutional Neural Network (DCNN) models and pooling operation for the classification of hematoxylin and eosin stain (H\&E) histological breast cancer images provided as a part of the International Conference on Image Analysis and Recognition (ICIAR) 2018 Grand Challenge on BreAst Cancer Histology (BACH) Images. Different data augmentation methods are applied to optimize the DCNN performance. We also investigated the efficacy of different stain normalization methods as a pre-processing step. The proposed network architecture using a pre-trained Xception model yields 92.50\% average classification accuracy.
\end{abstract}


\section{Introduction}
\begin{figure}[h]
\centering
\begin{subfigure}{0.2\textwidth}
	\includegraphics[width=\linewidth]{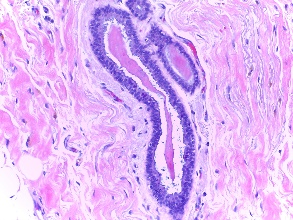}
	\caption{}
	\label{fig:1}
\end{subfigure}\hfil
\begin{subfigure}{0.2\textwidth}
	\includegraphics[width=\linewidth]{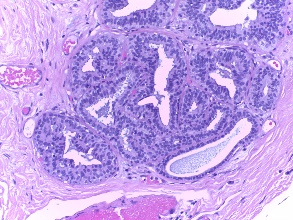}
	\caption{}
	\label{fig:2}
\end{subfigure}\hfil
\medskip
\begin{subfigure}{0.2\textwidth}
	\includegraphics[width=\linewidth]{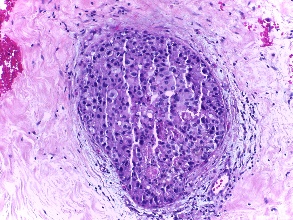}
	\caption{}
	\label{fig:4}
\end{subfigure}\hfil
\begin{subfigure}{0.2\textwidth}
	\includegraphics[width=\linewidth]{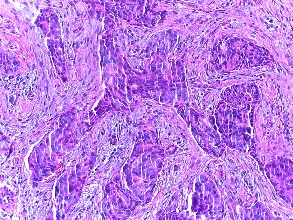}
	\caption{}
	\label{fig:5}
\end{subfigure}\hfil
\caption{Examples of H\&E stained breast histology microscopy images of (a) Normal, (b) Benign, (c) In situ, and (d) Invasive cases from ICIAR 2018 grand challenge on BreAst Cancer Histology (BACH) dataset}
\label{fig:ICIARExamples}
\end{figure}

Breast cancer is the most common form of cancer in women aged 20–59 years worldwide. According to the data provided by the American Cancer Society~\cite{BreastCancerStatistics}, in 2019, about 268,600 new cases of invasive breast cancer and about 62,930 new cases of in situ breast cancer will be diagnosed in which nearly 41,760 women will die from breast cancer. Tumors  can be subdivided into malignant (cancerous) and benign (non-cancerous) types, based on a variety of cell characteristics. Malignant tumors can be further categorized as being in situ (remain in place) or invasive. In situ carcinomas can form in the ducts or lobules of the breast and are not considerate to be invasive, but if left untreated, could increase the risk of developing invasive breast cancer~\cite{fondon2018automatic}. Early detection of the breast cancer is therefore important for increasing the survival rates of patients. The high morbidity and considerable healthcare cost associated with cancer has inspired researchers to develop more accurate models for cancer detection. Over the last five years, data mining and machine learning models have been used in a variety of research areas to dramatically improve our ability to discover emergent patterns within large datasets~\cite{DEABank,AMIN201982,REN201999,ZAZZARO2019100048,kassani2016pseudoinverse, Mardanisamani2019}. Developing computer-aided diagnosis (CAD) systems, integrated with medical image computing and machine learning methods, has become one of the major research paradigms for life-critical diagnosis~\cite{tosta2018computational}. CAD systems have been widely used in different fields, including mass detection~\cite{al2018fully}, lung cancer screening~\cite{khosravan2019collaborative}, mammography and breast histopathology image analysis~\cite{hamidinekoo2018deep}, medical ultrasound analysis~\cite{liu2019deep}, etc. Fig~\ref{fig:ICIARExamples} demonstrates some examples of breast histopathology images from the BACH dataset.
\subsection{Related studies}
In~\cite{rakhlin2018deep}, Rakhlin  et al., proposed a deep learning-based method for classification of H\&E stained breast tissue images. For each image, 20 crops of 400×400 pixels and 650×650 were extracted. Then, pre-trained ResNet-50, InceptionV3 and VGG-16 networks were used as feature extractors. Extracted features were combined through 3-norm pooling into a single feature vector. A LightGBM classifier with 10-fold cross-validation was used to classify extracted deep features. That method achieved an average accuracy 87.2 ± 2.6\% across all folds for classification of the breast cancer histology images.

In another study by Kwok~\cite{kwok2018multiclass}, four DCNN architectures i.e. VGG19, InceptionV3, InceptionV4 and InceptionResnetV2 were employed for the classification of H\&E stained histological breast cancer images for both multi-class and binary classification. In that study, 5600 patches with the size of 1495×1495 and stride of 99 pixels are extracted from the images. Different data augmentation methods were also employed to improve the accuracy of the method. In Kwok’s study, InceptionResnetV2 achieved the highest accuracy of 79\% for multi-class and 91\% for binary classification. 

Vang et al.~\cite{vang2018deep} proposed an ensemble-based InceptionV3 architecture for multi-class breast cancer image classification. Their proposed ensemble classifier included; majority voting, gradient boosting machine (GBM), and logistic regression to obtain the final image-wise prediction. The Vahadane~\cite{vahadane2015structure} stain-normalization method was utilized to normalize the stain images and with refinement achieved 87.50\% accuracy.

Another research study conducted by Sarmiento et al.~\cite{sarmiento2018automatic} proposed a machine learning approach using feature vectors extracted from different characteristics such as shape, color, texture and nuclei from each image. A Support Vector Machine (SVM) classifier with a quadratic kernel with 10-fold cross-validation was used to classify images but only achieved an overall accuracy of 79.2\%.

Finally, Nawaz et al.~\cite{nawaz2018classification} employed a fine-tuned AlexNet architecture for automatic breast cancer classification. The patches with the size of 512×512 pixels from training images were extracted and achieved an overall accuracy of 75.73\% for the patch-wise dataset and 81.25\% for the image-wise dataset.

The rest of the paper is organized as follows. Motivation and contributions are explained in next subsection. Section II provides a detailed description of materials and the proposed approach. Section III presents the experimental results obtained from proposed network architecture. Finally, the paper concludes in Section IV and provides future directions.

\subsection{Motivations and contributions}
Deep Convolution Neural Network (DCNN) models have achieved promising results in various medical imaging applications~\cite{al2018fully, khosravan2019collaborative, hamidinekoo2018deep, liu2019deep}. 
Moreover, it has been shown that data augmentation and stain normalization pre-processing steps are useful to get a more robust and accurate performance~\cite{kassani2019comparative, sajjad2019multi, ciompi2017importance}. These studies motivated us to explore the performance of different well-established DCNNs and also to verify the effectiveness of pre-processing and data augmentation techniques for breast cancer assessment from histopathological images. In the following, we list the main contributions of this study: 
\begin{enumerate}
	\item We demonstrate a new strategy for extracting bottleneck features from breast histological images using modified well-established DCNN networks. To learn more discriminative feature maps, we combine features extracted from convolutional layers after max pooling layers into a single feature vector, then, we used the obtained features as input to train a multilayer perceptron with 256 hidden neurons to classify the breast cancer histopathology images into four classes.

	\item For further improvement of the classification performance, pre-processing steps using different stain-normalization methods are employed. These preprocessing steps help to reduce the color inconsistency and therefore lead to improved efficiency in learning high-level features.
	
	\item The dataset provided for this study is very small. To increase the dataset size and improve the performance of our model, we utilized different data augmentation techniques such as horizontal and vertical flips, rotation, random contrast and random brightness.
\end{enumerate}
\begin{figure}[b]
	\centering
	\includegraphics[width=0.46\textwidth]{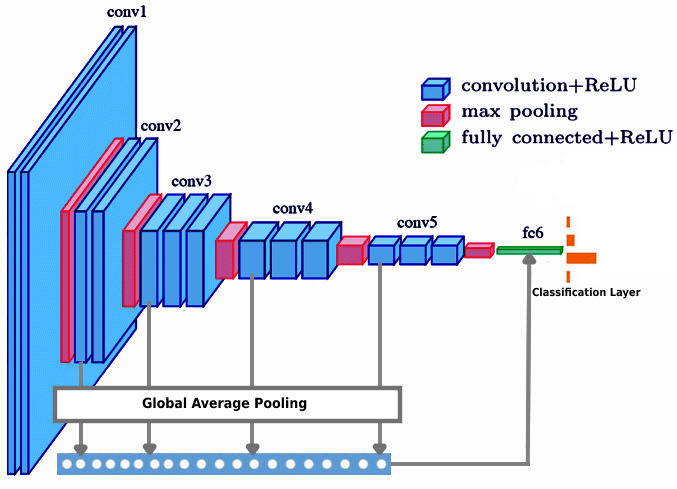}
	\caption{Illustration of our Proposed Classification Architecture.}
	\label{fig:ProposedArchitecture}
\end{figure}
\section{Methodology}
In this section, the proposed method based on DCNN architecture for training and predicting of breast cancer is explained. In the first step, stained histological breast cancer tissue images are pre-processed using stain normalization techniques. In the second step, data augmentation procedures are performed to address the issue of limited size of dataset and optimize the performance of DCNN models. In the third step, high-level features are extracted from pre-processed images using proposed network architecture from well-established DCNN models. Next, these extracted features are used as an input to a standard multilayer perceptron classifier. Finally, the performance of the proposed model is evaluated and reported on test images.
\subsection{Network architecture}\label{sec:NetworkArchitecture}
Feature extraction using DCNN models has achieved promising results in extracting high-level features for different classification tasks~\cite{ZHANG201910, LI2019347, CHOI2019259}. Since fine-tuning of well-established DCNN architectures has not previously achieved good performance on this dataset, for this study, we employ the DCNN descriptor approach~\cite{hall2018rapidly, gialampoukidis2019probabilistic, dong2015multi} to extract features in order to represent the discriminative characteristics of different classes sufficiently. In the proposed approach, features are extracted from the convolutional layer immediately after the max pooling layer and then followed by a global average pooling layer. Afterwards the extracted features are fused into a single feature vector. The extracted features, then are fed into a multilayer perceptron classifier for the  prediction. Fig~\ref{fig:ProposedArchitecture} illustrates the proposed architecture for breast cancer classification. For example, as demonstrated in Fig~\ref{fig:ProposedArchitecture}, we extract features from layers of 4, 7, 11 and 15 of VGG16 architecture, then apply a global average pooling to the extracted features, and next, we fuse them together to produce the final feature vector.

\subsection{Data pre-processing}\label{sec:StainNormalization}
\begin{figure}[hb]
	\centering
	\begin{subfigure}{0.15\textwidth}
		\includegraphics[width=\linewidth]{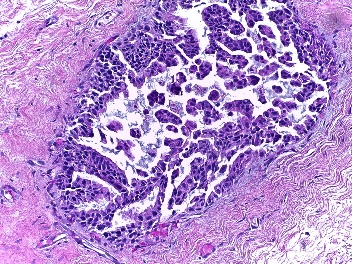}
		\caption{}
		\label{fig:6}
	\end{subfigure}\hfil
	\begin{subfigure}{0.15\textwidth}
		\includegraphics[width=\linewidth]{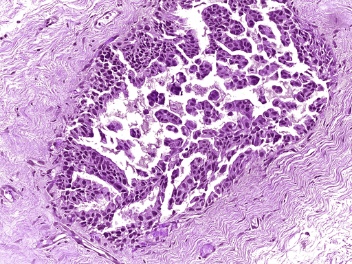}
		\caption{}
		\label{fig:7}
	\end{subfigure}\hfil
	\begin{subfigure}{0.15\textwidth}
	\includegraphics[width=\linewidth]{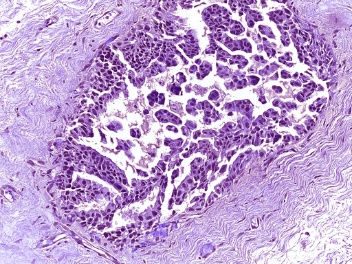}
	\caption{}
	\label{fig:8}
	\end{subfigure}\hfil
	
	\caption{Output of stain normalization methods: a) Original H\&E stained image, b) Macenko-normalized image, c) Reinhard-normalized image. }
	\label{fig:StainNormalization}
\end{figure}

Standardization of the H\&E stained images is an essential step before feeding the images to the deep networks. Therefore, in the first step, we stain normalize all histopathological images to reduce the color variation and hence have a better color consistency. We investigate the effectiveness of popular stain normalization techniques including methods proposed by Macenko et al.~\cite{macenko2009method}, and Reinhard et al.~\cite{reinhard2001color}. The original and stain normalized images are shown in Fig~\ref{fig:StainNormalization}. 

Before feeding images into DCNN architectures, we also need to apply another normalization method by subtracting the mean RGB value of ImageNet dataset images from all images of the training and test dataset~\cite{21-Yu2017}. The ImageNet mean RGB value is a precomputed constant derived from the ImageNet database~\cite{22-Krizhevsky2012}. 
\subsection{Data augmentation}\label{sec:DataAugmentation}
\begin{figure}[t]
	\centering
	\begin{subfigure}{0.15\textwidth}
		\includegraphics[width=\linewidth]{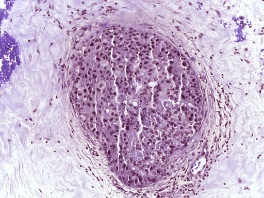}
		\caption{}
		\label{fig:9}
	\end{subfigure}\hfil
	\begin{subfigure}{0.15\textwidth}
		\includegraphics[width=\linewidth]{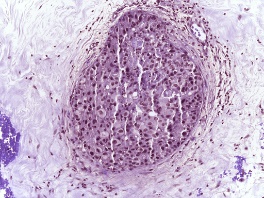}
		\caption{}
		\label{fig:10}
	\end{subfigure}\hfil
	\begin{subfigure}{0.15\textwidth}
	\includegraphics[width=\linewidth]{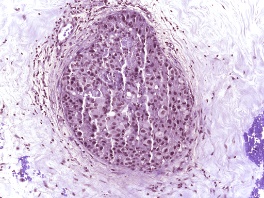}
	\caption{}
	\label{fig:11}
\end{subfigure}\hfil

	\medskip
	\begin{subfigure}{0.15\textwidth}
		\includegraphics[width=\linewidth]{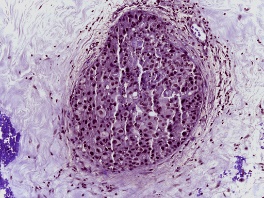}
		\caption{}
		\label{fig:12}
	\end{subfigure}\hfil
	\begin{subfigure}{0.15\textwidth}
		\includegraphics[width=\linewidth]{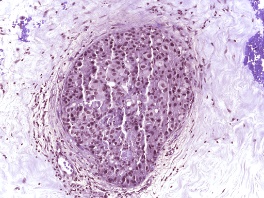}
		\caption{}
		\label{fig:13}
	\end{subfigure}\hfil
	\begin{subfigure}{0.15\textwidth}
	\includegraphics[width=\linewidth]{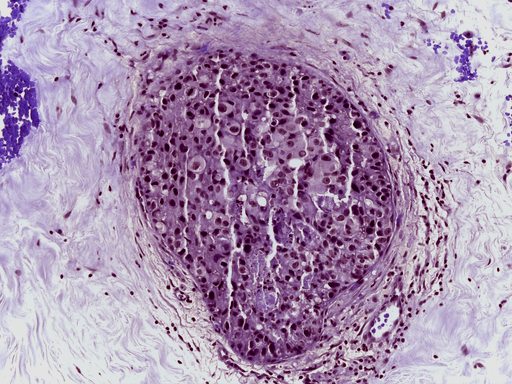}
	\caption{}
	\label{fig:14}
	\end{subfigure}\hfil
	
	\caption{The result of applying data augmentation techniques (a) an original image, (b) vertical flip (c) contrast adjustments and rotation (d) vertical flip, contrast adjustments and brightness correction (e) vertical flip, contrast adjustments and rotation, (f) contrast adjustments and brightness correction.}
	\label{fig:DataAugmentation}
\end{figure}

The performance of the DCNN predictive models may degrade due to the small size of training dataset. In this regards, different data augmentation techniques such as horizontal and vertical flips, rotation, contrast adjustments and brightness correction are applied to enlarge the dataset and improve the classification performance. Some examples of in situ cases after pre-processing and data augmentation steps are shown in Fig~\ref{fig:DataAugmentation}.

\subsection{Pre-trained DCNN feature extractors}
In this study, five DCNN architectures are employed as feature extractors, namely InceptionV3, InceptionResNetV2, Xception and two VGGNet models. Transfer learning is a method widely used in different tasks. In this method a large dataset from a source task is employed for training of a target task using the weights trained by the images from source dataset. The main advantage of transfer learning is the improvement of classifier accuracy and the acceleration of the learning process~\cite{KHAN20191}. Previous studies in the literature have demonstrated that transfer learning also has the potential to reduce the problem of overfitting~\cite{Lei2018}~\cite{LU201941}. Although the dataset is not the same, low-level features from the source dataset are generic features e.g. edges, contours and curves which are similar to the low-level features of target dataset~\cite{MEHDIPOURGHAZI2017228}.

\section{Experiment and Results}
\subsection{Dataset description}
The dataset used for this research is the ICIAR 2018 Grand Challenge ~\cite{ARESTA2019122} on BreAst Cancer Histology (BACH) Images publicly available at~\cite{ICIARDataset}. The goal of this challenge is to develop computer analysis systems that assist pathologists for accurate breast cancer assessment from histopathological images. The dataset consists of 400 H\&E stained histological breast tissue images with four categories namely as benign, normal, in-situ and invasive carcinoma evenly distributed (100 images per class). All images stored in tagged image file format (TIFF) with a magnification factor of 200× and a pixel size 0.42 µm * 0.42 µm. All images have the consistent shape of 2048 × 1536 pixels. We randomly divide the dataset into two parts, 300 images are used for training and 100 images for test data. In order to increase the size of the training dataset, we applied different data augmentation. The class distributions of dataset before and after data augmentation is presented in Table~\ref{tab:DatasetDistribution}. 
{\renewcommand{\arraystretch}{1.7}
	\begin{table}[h]
		\centering
		\scriptsize
		\caption{Total number of class distributions before and after data augmentation.}\label{tab:DatasetDistribution}
		
		\begin{tabular}{|l|l|l|l|l|l|}
			\hline
			\multicolumn{6}{|c|}{Number of images for each class} \\ \hline
			& Normal & Benign & In situ & Invasive & Total \\ \hline
			Original training Data & 75 & 75 & 75 & 75 & 300 \\ \hline
			Augmented training Data & 1155 & 1155 & 1155 & 1155 & 4620 \\ \hline
			Original test data & 25 & 25 & 25 & 25 & 100 \\ \hline
		\end{tabular}
	\end{table}
\subsection{Experimental Setup}\label{AA}
We do not extract patches for this experiment, unlike the majority of previous studies~\cite{chekkoury2012automated}\cite{Kamyar2018}. All images are downsized into 512×512 pixels using bicubic interpolation and normalized by subtracting the mean image computed from the training set. A fully connected layer trained with ReLU activation function and followed by a dropout~\cite{srivastava2014dropout} with a rate of 0.5 to prevent overfitting. $\beta 1 $, $\beta 2$ and learning rate for Adam optimizer were set to 0.6, 0.8 and 0.001 respectively. Weights are initialized from weights trained on ImageNet, as suggested by~\cite{ImageNetWeights} for all DCNNs. The batch size is set to 32, and we set 1000 epochs to train all models. Our experiment is implemented in Python using Keras package with Tensorflow as deep learning framework backend and run on Nvidia GeForce GTX 1080 Ti GPU with 11 GB RAM. For the proposed network architectures, descriptor features are extracted from blocks presented in Table~\ref{tab:ExtractedBlocks} for each pre-trained DCNN model. 
{\renewcommand{\arraystretch}{1.5}
\begin{table}[h]
	\centering
	\caption{Features are extracted from specific blocks for each pre-trained DCNN model.}\label{tab:ExtractedBlocks}
	\arrayrulecolor[rgb]{0.498,0.498,0.498}
	\begin{tabular}{|p{\mylength} | l|p{\mylength}}
		\arrayrulecolor{black}\hline
		Method            & Blocks                                                                                                           \\ 
		\arrayrulecolor[rgb]{0.498,0.498,0.498}\hline
		InceptionV3       & \begin{tabular}[c]{@{}l@{}} Blocks (11, 18, 28, 51, 74, 101, 120, \\152, 184, 216, 249, 263, 294) \end{tabular}  \\ 
		\hline
		InceptionResNetV2 & Blocks (11, 18, 275, 618)                                                                                        \\ 
		\arrayrulecolor{black}\hline
		Xception          & Blocks (26, 36, 126)                                                                                             \\ 
		\arrayrulecolor[rgb]{0.498,0.498,0.498}\hline
		VGG16             & Blocks (4, 11, 15)                                                                                               \\ 
		\arrayrulecolor{black}\hline
		VGG19             & Blocks (4, 7, 17)                                                                                                \\
		\arrayrulecolor[rgb]{0.498,0.498,0.498}\hline
	\end{tabular}
	\arrayrulecolor{black}
\end{table}
\subsection{Results and discussion}
The proposed framework is trained on five DCNN architectures, i.e. InceptionV3, InceptionResNetV2, Xception and two VGGNet models. The obtained results are compared with different existing stain-normalization techniques. We started our experiments by examining the effect of the stain normalization on performance. The performance of all architectures are evaluated based on the overall prediction accuracy. The obtained results of the plain architectures are summarized in Table~\ref{tab:ResultsPlain}. As shown in this Table, the Xception and InceptionV3 architecture individually give better average classification accuracy of 88.50\%, and 84.50\%, respectively.
{\renewcommand{\arraystretch}{1.5}
\begin{table}

	\centering
	\caption{Comparative analysis of the DCNN architectures on different stain normalization techniques. Bold value indicate the best result; underlined value represent the second-best result of the respective category.}\label{tab:ResultsPlain}
	\arrayrulecolor{black}
	
	\begin{tabular}{|l|l|l|l|}
		\hline
		~                       & Macenko & Reinhard & \begin{tabular}[c]{@{}l@{}} Average Accuracy \end{tabular}  \\ 
		\hline
		Plain-InceptionV3       & 84.00\% & 85.00\%  &  \uline{84.50\% }                                                             \\ 
		\hline
		Plain-InceptionResNetV2 & 85.00\% & 83.00\%  & 84.00\%                                                                       \\ 
		\hline
		Plain-Xception          & 87.00\% & 90.00\%  & \textbf{ 88.50\% }                                                            \\ 
		\hline
		Plain-VGG16             & 80.00\% & 81.00\%  & 80.50\%                                                                       \\ 
		\hline
		Plain-VGG19             & 77.00\% & 78.00\%  & 77.50\%                                                                       \\
		\hline
	\end{tabular}
	\arrayrulecolor{black}
\end{table}

\begin{table}
	\centering
	\caption{Evaluation results obtained from proposed network architectures. An asterisk beside the model name indicates  a modified DCNN architecture. The best result is shown boldface. Underlined value represent the second-best result of the respective category.}\label{tab:ResultsProposed}
	\arrayrulecolor{black}
	\begin{tabular}{|l|l|l|l|}

		\hline
		~                  & Macenko & Reinhard & \begin{tabular}[c]{@{}l@{}} Average Accuracy \end{tabular}  \\ 
		\hline
		InceptionV3*       & 90.00\% & 90.00\%  & \uline{ 90.00\% }                                                             \\ 
		\hline
		InceptionResNetV2* & 90.00\% & 88.00\%  & 89.00\%                                                                       \\ 
		\hline
		Xception*          & 91.00\% & 94.00\%  &  \textbf{92.50\% }                                                            \\ 
		\hline
		VGG16*             & 83.00\% & 87.00\%  & 85.00\%                                                                       \\ 
		\hline
		VGG19*             & 80.00\% & 84.00\%  & 82.00\%                                                                       \\
		\hline
	\end{tabular}
	\arrayrulecolor{black}
\end{table}

Table~\ref{tab:ResultsProposed} shows the obtained results of the proposed network architectures as well as average classification accuracies. The asterisk (*) indicates that the DCNN models are modified based on the proposed network architecture. As the results shown in Tables~\ref{tab:ResultsPlain} and~\ref{tab:ResultsProposed} of our preliminary analysis suggest, the Reinhard stain-normalization technique could achieve better classification accuracy than Macenko stain-normalization technique in most of the architectures. As shown in Table~\ref{tab:ResultsProposed}, the Xception* and InceptionV3* architectures individually gives better average classification accuracy of 92.50\%, and 90.00\%, respectively. Xception* architecture has 92.50\% average accuracy while VGG19* and VGG16* have average accuracies of 82.00\% and 85.00\%, respectively. This means the gap in accuracy is 10.50\% and 7.50\%, respectively in favor of Xception*. The gap of Xception* compared to InceptionResNetV2* and InceptionV3*, is 3.50\% and 2.50\%, respectively. So, Xception* has the best average accuracy and the VGG19* has the worst accuracy among all counterparts. Similar conclusions can be drawn for other models. It is also inferred from Table~\ref{tab:ResultsProposed} that employing the Reinhard stain-normalization method tends to give an improvement of overall accuracy by high margin of 3.00\% compared to the Macenko stain-normalization method. The Xception* architecture proved to be most effective at classifying examples belonging to the Normal and Invasive classes using Macenko stain-normalization method and Benign and Invasive classes using Reinhard stain-normalization method as illustrated in the form of confusion matrices in Fig~\ref{fig:ConfusionMatrices}.
\begin{figure}[h]
	\centering
	\begin{subfigure}{0.24\textwidth}
		\includegraphics[width=\linewidth]{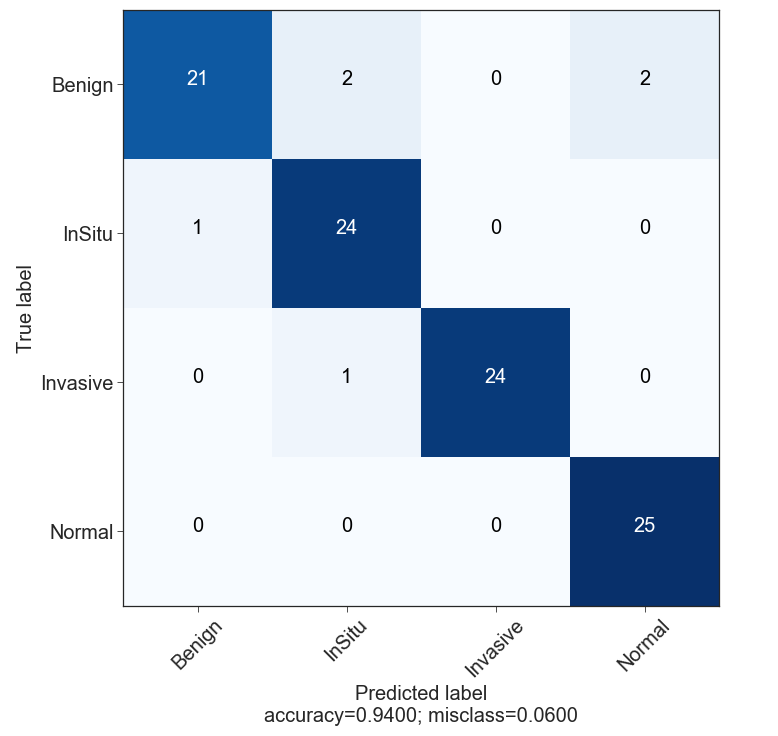}
		\caption{}
		\label{fig:14}
	\end{subfigure}\hfil
	\begin{subfigure}{0.24\textwidth}
		\includegraphics[width=\linewidth]{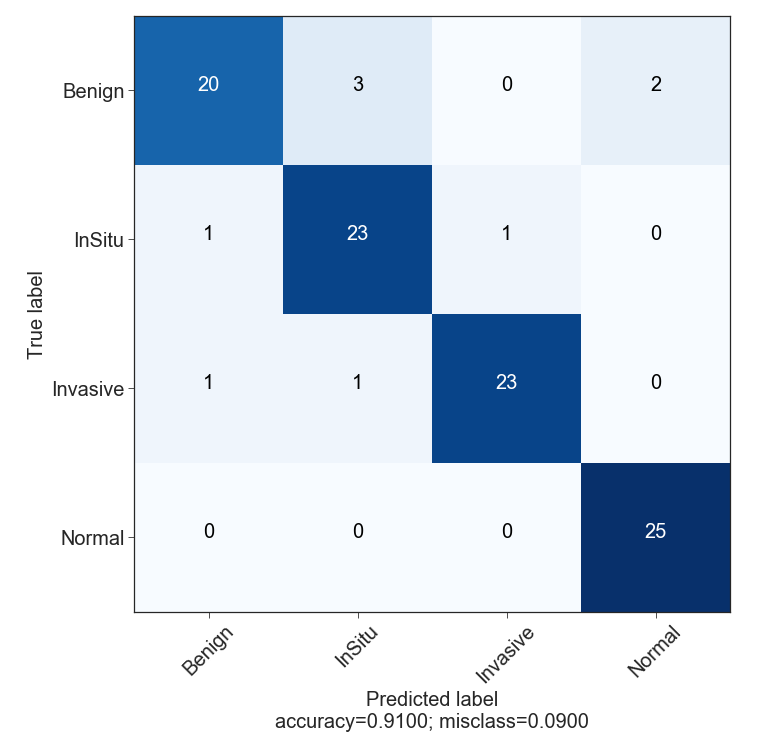}
		\caption{}
		\label{fig:15}
	\end{subfigure}\hfil
	
	\caption{(a) Confusion matrix of breast cancer classification using Xception* model and Reinhard et al. stain-normalization. (b) Confusion matrix of breast cancer classification using Xception* model and Macenko et al. stain-normalization.}
	\label{fig:ConfusionMatrices}
\end{figure}

\subsection{Comparative analysis of accuracy with other methods}
For evaluating the effectiveness of the proposed method, a comparative analysis with the results of some of the previously published work from the same dataset is presented in Table~\ref{tab:ComparativeAnalysis}. It can be observed from Table~\ref{tab:ComparativeAnalysis} that the methods in~\cite{vang2018deep}, ~\cite{rakhlin2018deep}, ~\cite{nawaz2018classification}, ~\cite{sarmiento2018automatic} and ~\cite{kwok2018multiclass} give an accuracy of 87.50\%, 87.20\%, 81.25\%,79.20\%, and 79.00\% respectively, whereas, the results obtained using the network architecture used here, give an accuracy of 92.50\%. These results confirm the superiority of our learner in terms of accuracy compared to other similar methods.
{\renewcommand{\arraystretch}{1.5}
\begin{table}[h]
	\centering
	\caption{Comparative analysis with other methods}\label{tab:ComparativeAnalysis}
	\arrayrulecolor{black}
	\begin{tabular}{|p{\mylength} | l|p{\mylength}}
		\hline
		Methods         & Accuracy  \\ 
		\hline
		Kwok~\cite{kwok2018multiclass}            & 79.00\%   \\		
		\hline
		Sarmiento et al~\cite{sarmiento2018automatic} & 79.20\%   \\ 
		\hline
		Nawaz et al~\cite{nawaz2018classification}     & 81.25\%   \\ 
		\hline
		Rakhlin et al~\cite{rakhlin2018deep}   & 87.20\%   \\ 
		\hline
		Vang et al~\cite{vang2018deep}      & 87.50\%   \\ 
		\hline
		\textbf{Proposed architecture}         & \textbf{92.50\%}  \\ 
		\hline
	\end{tabular}
	\arrayrulecolor{black}
\end{table}

\section{Conclusion}
In this paper, we proposed an effective deep learning-based method using a DCNN descriptor and pooling operation for the classification of breast cancer. We also employed different data augmentation techniques to boost the performance of classification. The effect of different stain normalization methods are also investigated. Experimental results demonstrate the proposed network architecture using pre-trained Xception model outperforms all other DCNN architectures with 92.50\% in terms of average classification accuracy. For future work, we aim to further improve the classification accuracy by utilizing deep learning-based ensemble models and better stain normalization techniques. 

\bibliographystyle{IEEEtran}
\bibliography{ref}

\end{document}